\RequirePackage{fix-cm}
\documentclass[twocolumn]{svjour3}          
\smartqed  
\usepackage{graphicx}
\usepackage{amssymb} 

\usepackage{subfigure}
\usepackage{amsmath}
\usepackage{amsfonts}
\usepackage{amsbsy}
\usepackage{times}
\usepackage{mathptmx}
\usepackage[left]{lineno}

\usepackage{color} 

 \journalname{Granular Matter}
\begin{document}

\title{Velocity fields and particle trajectories for bed load over subaqueous barchan dunes\thanks{This is a pre-print of an article published in Granular Matter, 21:75, 2019. The final authenticated version is available online at: http://dx.doi.org/10.1007/s10035-019-0928-0}}


\author{Jo\~ao Luiz Wenzel \and
				Erick de Moraes Franklin 
}


\institute{Jo\~ao Luiz Wenzel \at
              School of Mechanical Engineering, UNICAMP - University of Campinas\\
							 \email{wenzel@fem.unicamp.br}               
									 \and
					 Erick de Moraes Franklin (corresponding author)\at
              School of Mechanical Engineering, UNICAMP - University of Campinas \\
              Tel.: +55-19-35213375\\							
              orcid.org/0000-0003-2754-596X\\
              \email{franklin@fem.unicamp.br}             
}

\date{Received: date / Accepted: date}

\maketitle

\begin{abstract}
This paper presents an experimental investigation of moving grains over subaqueous barchan dunes that consisted of spherical glass beads of known granulometry. Prior to each test run, a pre-determined quantity of grains was poured inside a closed conduit, and the grains settled on its bottom wall forming one conical heap. As different turbulent water flows were imposed, each heap evolved to a barchan dune, which was filmed with a high-speed camera. An image processing code was written to identify some of the moving grains and compute their velocity fields and trajectories. Our results show that the velocity of grains varies along the barchan dune, with higher velocities occurring close to the dune centroid, and that grains trajectories are intermittent. Depending on the region over the dune, we found that the velocity fields present values within 1 and 10\% of the cross-sectional mean velocity of the fluid. Considering the average trajectories of grains moving over a given dune, their mean displacement varies within 30 and 60 grain diameters and their characteristic velocities within 10 and 20 \% of the cross-sectional mean velocity of the fluid. The displacement time varies between 30 and 90 \% of the settling time, and it seems to have two asymptotic behaviors: one close to bed load inception and other far from it. When compared with bed load over a plane bed, we observe that grains have the same mean velocity, but they travel distances up to 5 times larger, with higher densities of moving grains.

\keywords{Barchan dunes \and bed load \and closed-conduit flow \and turbulent flow}
\end{abstract}

\section{Introduction}
 
The interaction between granular matter and fluids is frequent in both nature and industry. One example commonly observed is the transport of sand by a fluid flow, such as happens in rivers, deserts, oceans, and pipelines. If sand is entrained in the horizontal (or nearly horizontal) direction as bed load, a moving layer that keeps contact with the static part of the bed, sand dunes usually grow \cite{Bagnold_1,Raudkivi_1,Yalin_1}. Their growth is the result of local erosion and deposition rates, where the perturbation of the fluid flow is known to be the unstable mechanism \cite{Andreotti_1,Andreotti_2,Elbelrhiti,Charru_3,Claudin_Andreotti}. Under one-directional flow and limited sand supply, sand dunes evolve to crescentic shape dunes, known as barchan dunes \cite{Herrmann_Sauermann,Sauermann_1}, or simply barchans, and their length depends on the length scale \cite{Hersen_1}:

\begin{equation}
L_{drag} = d \frac{\rho_s}{\rho},
\label{ldrag_equation}
\end{equation}

\noindent where $d$ is the diameter of grains, and $\rho_s$ and $\rho$ are the densities of the granular material and fluid, respectively. Therefore, the length and time scales of barchans go from 100 m and 1 year on deserts down to 10 cm and 1 minute for subaqueous barchans \cite{Claudin_Andreotti}.

Because barchan dunes are formed when grains are entrained as bed load, with the moving layer exchanging mass with the fixed part of the bed, the distribution of moving grains and their trajectories are important to understand the morphology and dynamics of barchans. For aeolian dunes, the moving grains effectuate ballistic flights over distances much larger than the grain diameter \cite{Bagnold_1}, while for subaqueous dunes the grains move mainly by rolling or sliding over each other, over distances comparable to the grain diameter \cite{Franklin_8}. If, on the one hand, there are models for the displacements of grains over aeolian and subaqueous dunes \cite{Sauermann_2,Sauermann_4,Kroy_C,Hersen_3,Pahtz_1}, on the other hand there are not, at present, precise measurements of the displacements of individual grains over barchans.

Few previous papers reported experiments on the displacements of individual grains within a bed-load layer under water flows. For the case of plane granular beds, Seizilles et al. \cite{Seizilles} investigated the displacements of individual grains under laminar flows, while Lajeunesse et al. \cite{Lajeunesse} and Penteado and Franklin \cite{Penteado} investigated them under turbulent flows. Lajeunesse et al. \cite{Lajeunesse} performed experiments on a water flume, where steady-state free-surface turbulent flows were imposed over different granular beds. The Reynolds numbers based on the water depth were within $1500$ and $6000$, and the beds consisted of quartz particles with median diameters of $1.15\,mm$, $2.24\,mm$, or $5.5\,mm$, corresponding to $12\,\leq\, Re_* \,\leq\,500$, where

\begin{equation}
Re_* = \frac{u_* d}{\nu}
\label{eq:Refric}
\end{equation}

\noindent is the Reynolds number at the grain scale. In Eq. \ref{eq:Refric}, $u_*$ is the shear velocity and $\nu$ the kinematic viscosity. Grains velocities, displacement lengths, and durations of flights were computed from the displacements of individual grains, which were obtained from images acquired with a high-speed camera. Concerning the grain velocities, Lajeunesse et al. \cite{Lajeunesse} found that the distributions of their transverse and longitudinal components follow Gaussian and decreasing exponential laws, respectively. In addition, they showed that the flight duration scales with the settling velocity of a single grain, the surface density of moving grains scales with $\theta - \theta_{c}$, and the grains velocity and flight length scale with $\theta^{1/2}-\theta_{c}^{1/2}$, where $\theta$ and $\theta_{c}$ are the Shields and the threshold Shields numbers. The Shields number is the ratio between the fluid drag and the particle friction (given by Coulomb's law), and typical values for bed load are within $0.01 < \theta < 1$. For turbulent flows, the Shields number is given by:

\begin{equation}
\theta = \frac{u_*^2}{\left( \rho_s / \rho -1 \right) g d},
\end{equation}

\noindent where $g$ is the magnitude of gravity. The threshold Shields number is the value of the Shields number at the inception of bed load.

Penteado and Franklin \cite{Penteado} performed their experiments in a closed-conduit channel of rectangular cross section. In the experiments, fully-developed turbulent flows were imposed over plane beds of known granulometry, and the granular bed was filmed with a high-speed camera. The granular beds consisted of glass spheres with density $\rho_s\,=\,2500\,kg/m^3$ and diameter ranging from $d\,=\,400\,\mu$m to $d\,=\,600\,\mu$m. The Reynolds numbers based on the hydraulic diameter (twice the channel height) were within $1.6 \times 10^4$ and $2.7 \times 10^4$, and $5\,\leq\, Re_* \,\leq\,11$. From the acquired images, Penteado and Franklin \cite{Penteado} computed the velocity fields and trajectories of individual grains. The authors found that the mean longitudinal displacement normalized by the grain diameter is of the order of 10, the  mean displacement velocity normalized by the shear rate and grain diameter is of the order of 0.1, and the mean displacement time normalized by the grain diameter and settling velocity is of the order of 10.

Numerical studies have been conducted since the 2000s to investigate the formation of single barchans as well as of barchan fields. The first studies made use of continuum equations adapted for granular media, such as, for instance, Sauermann et al. \cite{Sauermann_4}, Kroy et al. \cite{Kroy_A,Kroy_C}, Hersen \cite{Hersen_3} and Schw{\"a}mmle and Herrmann \cite{Schwammle}. In those models, information such as the characteristic lengths and times are essential to fit adjustable constants. More recently, Lagrangian methods, such as the 3D cellular automaton model \cite{Zhang_2}, and Eulerian-Lagrangian methods, such as CFD-DEM (Computational Fluid Dynamics - Discrete Element Method) \cite{Khosronejad}, are being used in the study of barchan dunes. One of their advantages is that information on the grain scale can be obtained even inside a barchan dune, which is considerably difficult to be obtained from experiments. However, the coupling between the fluid flow and bed load rests to be validated in those simulations, and, therefore, experimental measurements at the grain scale are still important.

Zhang et al. \cite{Zhang_2} investigated numerically the mean residence time of individual grains in barchans in order to understand the contribution of each grain for the existence of a barchan dune. Using a 3D cellular automaton model, the authors tracked individual particles from the time they are incorporated to the barchan dune to the time they leave it by one of its horns. For different size and flow conditions, they showed the existence of two antagonistic lateral fluxes of grains, which are outward on the windward side (upstream the dune crest) and inward on the lee face (downstream the dune crest). Because of that antagonistic movement, grains in the central region tend to remain in this region (and in the dune), while grains close to lateral flanks tend to move to the horns and leave the dune. As a result, Zhang et al. \cite{Zhang_2} found that the mean residence time does not depend strongly on the dune size. Instead, it is given by the surface of the longitudinal central slice of the dune divided by the input sand flux.

\begin{sloppypar}
Recent studies investigated the formation of subaqueous barchans from initially conical heaps by considering the growth of horns \cite{Alvarez,Alvarez3}. Alvarez and Franklin \cite{Alvarez} measured experimentally the growth of horns on conical heaps under turbulent water flows. For the length of horns as a function of time, Alvarez and Franklin \cite{Alvarez} showed the existence of an initially positive slope, corresponding to its development, and a final plateau, corresponding to an equilibrium length for horns. Based on the evolution of horns, they proposed the characteristic times $0.5 t_c$ for the growth and $2.5 t_c$ for equilibrium of barchans, where $t_c$ is a characteristic time for the displacement of barchans computed as the length of the bedform divided by its celerity. Alvarez and Franklin \cite{Alvarez3} investigated the trajectories of individual grains during the formation of horns. The authors showed that most of grains forming the horns during their growth migrate from upstream regions on the periphery of the initial heap. In addition, they showed that individual grains have transverse displacements by rolling and sliding that are not negligible, contrasting with the general assertions for aeolian dunes that transverse displacements are due mainly to the diffusive effect of reptons and that horns grow mostly with grains originally in the flanks of the initial pile.
\end{sloppypar}

\begin{sloppypar}
The velocity fields and trajectories of grains over barchan dunes are important to understand not only the growth of these forms, but also their morphology and migration. In addition, this information is essential to fit adjustable constants in continuum models and to validate Lagrangian and Eulerian-Lagrangian methods. However, there are very few experimental studies on displacements of individual grains over subaqueous forms \cite{Alvarez,Alvarez3}. Because the water flow varies over the dune surface, the results obtained for bed load over plane granular beds are not valid for bed load over barchans. To the authors' knowledge, no previous study reported the velocity fields of grains over developed barchan dunes.
\end{sloppypar}

This paper presents an experimental investigation on the displacements of individual grains over subaqueous barchans. The experiments were performed in a closed conduit of transparent material and the dunes were filmed with a high-speed camera. The velocity fields and the trajectories of individual grains were computed from the acquired images with numerical scripts written in the course of this work. Our results show that the fields of grain velocities present local values within 1 and 10\% of the cross-sectional mean velocity of the fluid, and that, considering the average trajectories of grains moving over a given dune, grains have mean displacements within 30 and 60 grain diameters, with characteristic velocities within 10 and 20 \% of the cross-sectional mean velocity of the fluid. The displacement time varies between 30 and 90 \% of the settling time, and it seems to have two asymptotic behaviors: one close to bed load inception and other far from it. When compared with bed load over a plane bed, we observe that grains have the same mean velocity, but they travel distances up to 5 times larger, with higher densities of moving grains. Those findings contribute to increase our understanding of the dynamics of barchan dunes and can represent significant advancements on future numerical simulations involving barchans.

The next sections present the experimental setup and experimental results. The following section presents the conclusions.

\section{Experimental setup}
\begin{figure}[ht]	
 	\centering
 	\includegraphics[width=0.9\columnwidth]{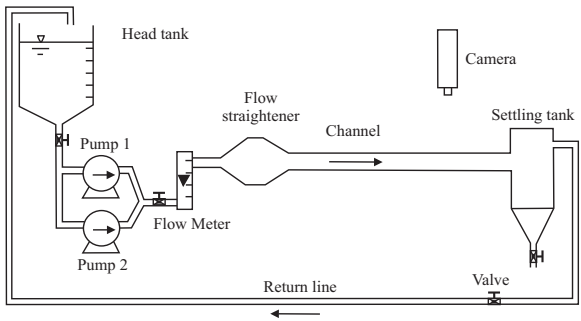}
 	\caption{Layout of the experimental setup.}
 	\label{fig:1}
\end{figure}

\begin{figure}[ht]	
 	\centering
 	\includegraphics[width=0.9\columnwidth]{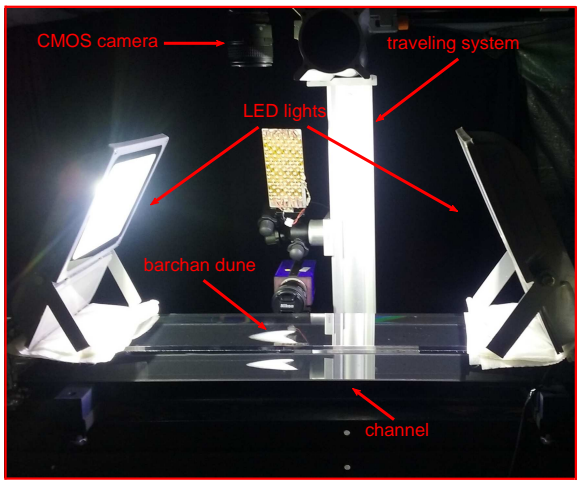}
 	\caption{Photograph of the test section.}
 	\label{fig:test_section}
\end{figure}

The experimental setup consisted of, in addition to a high speed camera and LED (Light-Emitting Diode) lamps, a water reservoir, centrifugal pumps, an electromagnetic flow meter, a flow straightener, a 5 m-long channel, and a settling tank, the water flowing in closed loop in the order just described. The channel was a closed conduit of rectangular cross section, 160 mm wide by 50 mm high, made of transparent material (plexiglass). Its downstream part contained the test section, which started 3 m downstream of the channel inlet and was 1 m long, the remaining 1 m section connecting the test section to the settling tank. The bottom wall of the test section was made of black plexiglass in order to minimize undesired reflection from that wall. The flow rates were adjusted with a set of valves, and the water flow was homogenized upstream of the channel entrance by the flow straightener, which consisted of a divergent-convergent nozzle filled with glass spheres. Fig. \ref{fig:1} shows a layout of the experimental setup and Fig. \ref{fig:test_section} presents a photograph of the test section.

Prior to each test, the grains were poured in the test section, which was previously filled with water, and they formed a single conical heap at the bottom wall of the channel. Afterward, for each test run, a turbulent water flow was imposed in the channel, deforming the initial pile into a barchan dune. With that procedure, each experiment concerned one single barchan that loosed grains by its horns, and, therefore, decreased slowly in size while migrating. The employed fluid was tap water at temperatures within 24 and 26 $^o$C, and the employed grains were round glass beads with $\rho_s = 2500$ kg/m$^3$ and bulk density of 1500 kg/m$^3$. The grains were separated in two different populations according to their diameter: $0.40$ mm $\leq\,d\,\leq$ $0.60$ mm and $0.15$ mm $\leq\,d\,\leq$ $0.25$ mm, called here types 1 and 2, respectively; therefore, $d_1$ = 0.50 mm and $d_2$ = 0.20 mm are the mid-range mean diameters of types 1 and 2, respectively. The cross-sectional mean velocities $U$ were 0.243, 0.295 and 0.365 m/s, which correspond to Reynolds numbers based on the hydraulic diameter $Re=\rho U D_h /\mu$ of $1.9 \times 10^4$, $2.2 \times 10^4 $ and $2.8 \times 10^4 $, respectively, where $\mu$ is the dynamic viscosity of the fluid and $D_h$ is the hydraulic diameter of the channel. The initial heaps were formed with either 6.2 g or 10.3 g of glass beads, corresponding to initial volumes of 4.1 and 6.9 cm$^3$, respectively. We employed white and black glass beads, where 98\% of grains were white and 2\% black, the latter being used as tracers. Tracers were used because they are easier to track along images and, as they had the same diameter and surface characteristics as the other grains, we assumed that their displacements characterize the whole moving bed.

A high-speed camera of CMOS (Complementary Metal Oxide Semiconductor) type was placed above the channel in order to record the bed evolution (Fig. \ref{fig:test_section}). We used either a camera with a spatial resolution of 1600 px $\times$ 2560 px at frequencies up to 1400 Hz or one with a resolution of 1280 px $\times$ 1024 px at frequencies up to 1000 Hz, both controlled by a computer. In our tests, we set the frequency to values within 200 Hz and 300 Hz depending on the average velocity of grains, and we used a Nikkor lens of $60\,mm$ focal distance and F2.8 maximum aperture. LED lamps were branched to a continuous current source in order to supply the required light and, at the same time, avoid beating between the camera and light frequencies. Prior to the beginning of tests, a calibration procedure which consisted of taking one picture from a scale placed in the water channel was performed, allowing the conversion from pixels to a physical system of units. For the lower resolution camera, we set the ROI (region of interest) to 800 px $\times$ 1024 px to better fit a field of view of 79.4 mm $\times$ 101.7 mm, whereas for the higher resolution one we set the ROI to 1208 px $\times$ 1648 px in order to fit a field of 84.3 mm $\times$ 115.0 mm; therefore, in the acquired images the areas covered by particles of type 1 are of the order of 25 and 40 px for the lower and higher resolution cameras, respectively, and of the order of 5 and 10 px in the case of type 2 particles. The images were recorded in RGB (red, blue and green) format. An example of movie, showing the motion of grains over a barchan dune, is available as Supplementary Material \cite{Supplemental}.

During the tests we did not impose an influx of grains. In this way, the barchan dune decreased in size while migrating along the channel. However, changes in both size and shape were negligible within the duration of each test (between 4 and 65 s).

\subsection{Image processing}
\label{section_image_processing}

The acquired images were processed by numerical scripts written in the course of this work, which use a PTV (Particle Tracking Velocimetry) approach that compares pairs of images, finds the tracers, and follow them along images. Once the tracers are identified in each movie frame, an Eulerian or a Lagrangian framework can be adopted to find the velocity fields or the trajectories of tracers, respectively.

The first step in the image processing code is to remove the image background. For that, the scripts convert the images to grayscale in order to better differentiate the black beads from the background, which is also black but with whiter shades than the black beads. With the background removed, the images are binarized, the main properties of the dune such as its centroid and area are obtained, and the tracers are identified. Given the presence of noise with size of 1 px, caused by small changes in light or small displacements (due to vibration) of the camera, the next step is the filtering with an area filter to distinguish the tracers from noise. This is necessary because the tracers are relatively small objects (covering from 5 to 40 px in the images). The area filter consists in the removal of objects with areas equal or smaller than 2 to 10 px, depending on the tested field. Once applied the filter, all the tracers are identified in each image.

\begin{figure}[h!]
\begin{center}
	\begin{tabular}{c}
	\includegraphics[width=0.90\columnwidth]{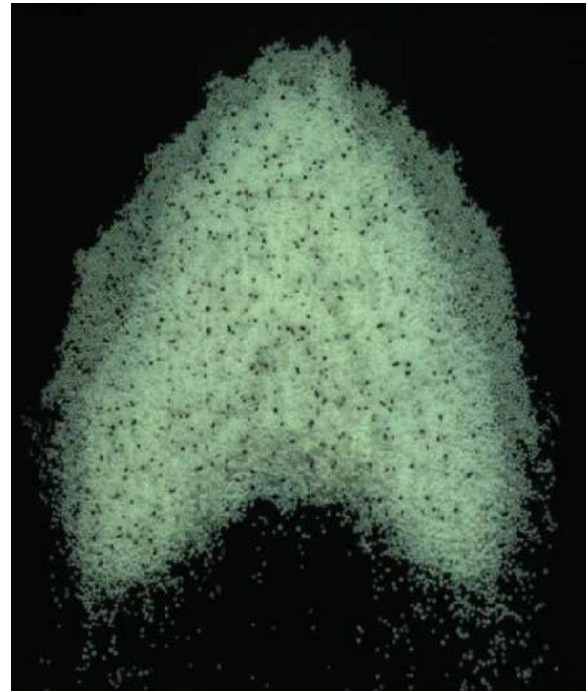}\\
	(a)\\
	\includegraphics[width=0.90\columnwidth]{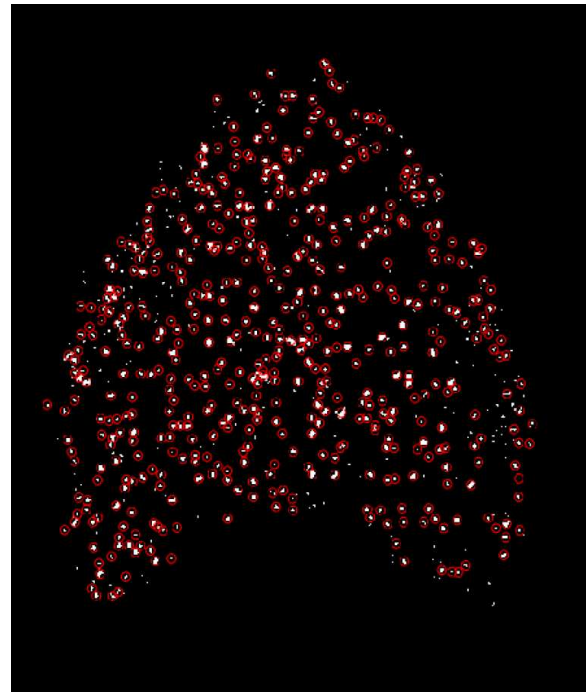}\\
	(b)
	\end{tabular} 
\end{center}
 	\caption{Top views of a barchan dune showing the visible black tracers: (a) raw image showing the tracers over the dune; (b) processed image with all the tracers identified by red circles. This dune was formed with $m$ = 10.3 g of type 1 grains ($d_1$ = 0.50 mm), under $U$ = 0.365 m/s. This corresponds to $Re_*$ = 10 and $\theta$ = 0.055. In the images, the flow is from top to bottom.}
 	\label{fig:tracers}
\end{figure}

Since the proportion of tracers is small (2\% of the dune), they can be tracked along images. For that, the code follows a sequence of rules to match the tracers in consecutive images. Those rules take advantage of previous knowledge concerning bed load under water flows, namely that grains move mainly in the longitudinal direction with velocities of the order of 10\%  of the cross-sectional mean velocity of water. Because the acquiring frequencies were sufficiently high, only four rules are necessary to match the tracers in image pairs: (i) to limit the maximum displacement of grains to within 30 and 50\% of the mean fluid displacement; (ii) to limit the variation of the area of a given tracer to 40 \%, i.e., a tracer in a given image must have an area within 60\% and 140\% of a tracer in the preceding image to allow matching; (iii) with the exception of tracers in the recirculation region, to assure that the longitudinal displacement is downstream; (iv) in case more than one tracer in one image can match a certain tracer in another image after the three previous rules were applied, to chose the tracer with the smallest transverse displacement between the considered images. Tracers that do not comply with those four rules are assumed to be no longer visible in the succeeding image, having left the dune by its horns or been buried under the bed, for instance. Once the tracers are identified and matched along images, they are labeled and their centroid positions stored together with the dune properties. Figs. \ref{fig:tracers}(a) and \ref{fig:tracers}(b) show, respectively, raw and processed images of the top view of a barchan dune. The black tracers over the surface of the barchan dune are visible in the raw image, and they are identified in the processed image by red circles with origin in the tracer centroids. 

The last steps are the computations in Eulerian and Lagrangian frameworks. For the Eulerian framework, the centroid positions of all matched tracers in consecutive images are subtracted and multiplied by the acquiring frequency. Next, Cartesian meshes with origin at the dune centroid are built, and the average velocity of grains in a cell is computed for each cell of the mesh. Finally, a temporal average is computed for each cell by considering all images of a test run (999 or 1299 images for tests with $Re$ = $2.8 \times 10^4 $ and 5997 for the remaining tests). With this procedure, fields of mean velocity of tracers are obtained for the considered mesh size. Sensitivity tests were performed in the meshes to determine the smallest size allowing the spatio-temporal averages. Based on that, we've chosen cell sizes of 25 px $\times$ 25 px.

For the Lagrangian framework, the evolution of centroid positions along time is determined for matched tracers, and the typical distances and velocities are computed.

\section{Results}

\subsection{Eulerian Framework}
 
Based on the centroid positions of tracers, Cartesian meshes were built, the mean velocities in each cell computed, and, afterward, a temporal average considering all images was computed for each cell, $<\overline{v}>$, as described in Subsection \ref{section_image_processing}. With this procedure we computed the mean velocity fields of tracers over a barchan dune. Because the only difference between tracers and the remaining grains was their color (they were both of colored glass), we assume in the following that the their behavior is representative of all grains forming the dune.

\begin{figure}[h!]
\begin{center}
	\includegraphics[width=0.90\columnwidth]{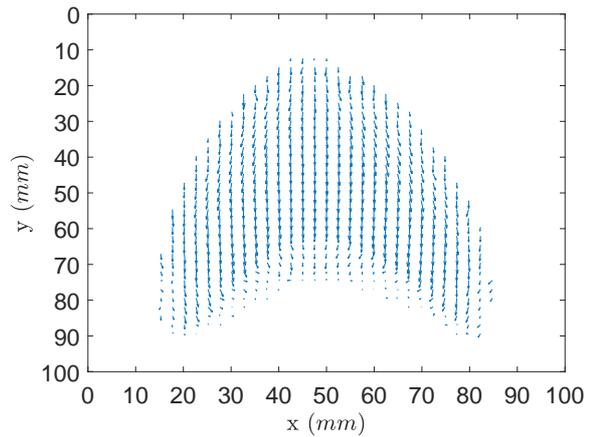}\\
\end{center}
 	\caption{Mean velocity field of grains over a dune formed with $m$ = 10.3 g of type 2 grains under $U$ = 0.295 m/s, corresponding to $Re$ = $2.2 \times 10^4$, $Re_*$ = 8 and $\theta$ = 0.038. The magnitude of arrows is proportional to that of velocities.}
 	\label{fig:quiver1}
\end{figure}

Figure \ref{fig:quiver1} presents the mean velocity field of grains over a barchan formed from 10.3 g of type 2 grains under a water flow with $U$ = 0.295 m/s. This corresponds to $Re$ = $2.2 \times 10^4$, $Re_*$ = 8 and $\theta$ = 0.038. In the figure, the abscissa and ordinate correspond, respectively, to the transverse ($x$) and longitudinal ($y$) coordinates, the magnitude of arrows is proportional to that of velocities, and the direction of arrows is the same as that of velocities.

\begin{sloppypar}
From Fig. \ref{fig:quiver1} we note that velocities increase from upstream position toward the dune centroid (and the dune crest), with a stronger longitudinal component than the transverse component, but with considerable transverse components close to the lateral flanks of the dune and in the region downstream of the crest. In this region, we observe small vectors pointing to the symmetry line of the barchan, corresponding to the mean velocities of grains falling by avalanches in the lee side. Just downstream of these vectors, we observe even smaller vectors with relatively strong transverse components pointing toward the symmetry line of the barchan, and which correspond to the mean velocities of grains in the recirculation region. Those regions and the motion of grains can be identified in the movie available as Supplementary Material \cite{Supplemental}. All the other test conditions showed similar behaviors, and their respective mean fields are also available as Supplementary Material \cite{Supplemental}.
\end{sloppypar}

An easier way to localize the mean grain velocities is by identifying the mean values of velocities by using polar coordinates centered at the dune centroid. The radial and angular plots were used by Alvarez and Franklin \cite{Alvarez3} to localize the origin of grains migrating to horns, and are used in the following to localize the magnitude of mean velocities.

\begin{figure}[h!]
\begin{center}
	\begin{tabular}{c}
	\includegraphics[width=0.90\columnwidth]{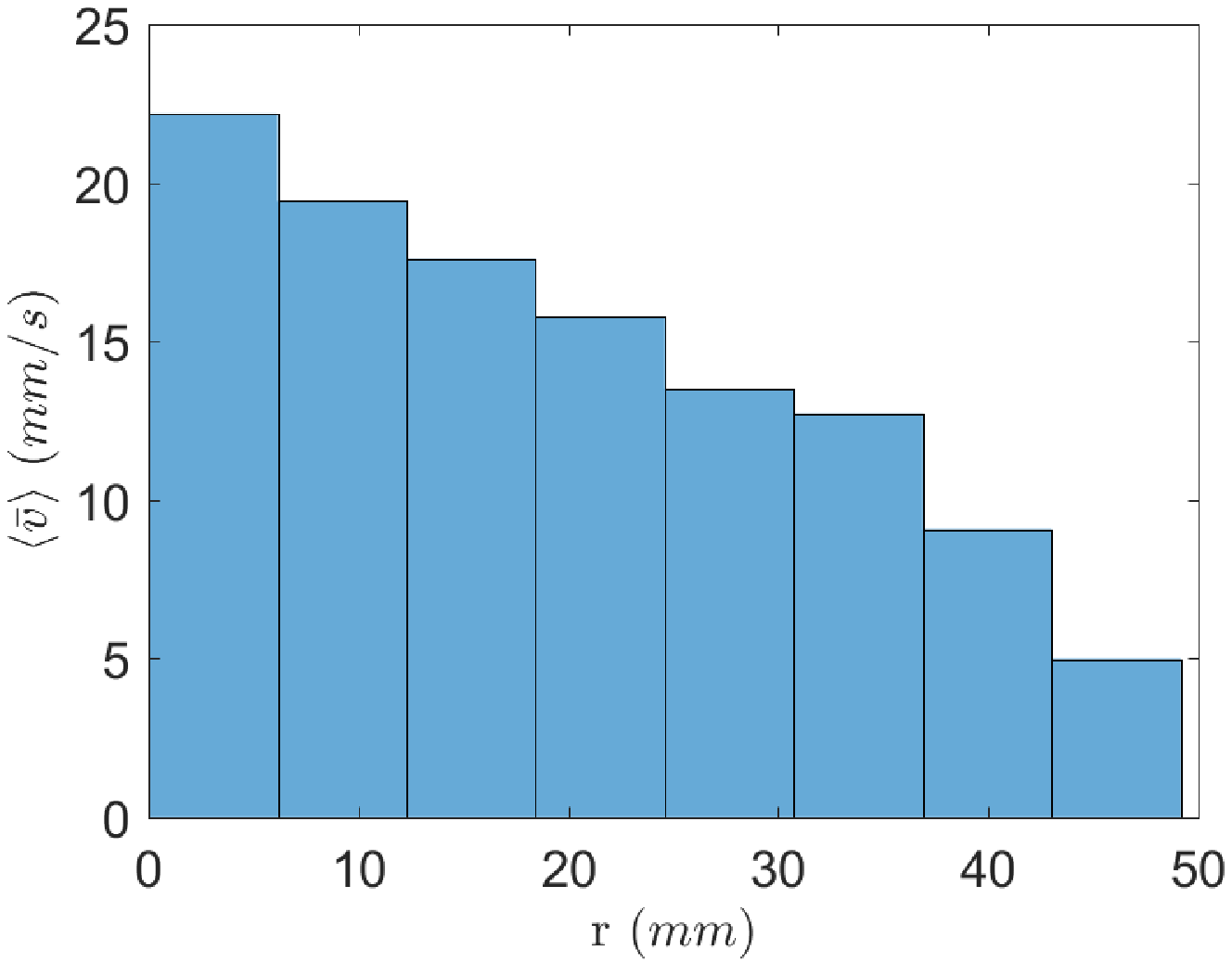}\\
	(a)\\
	\includegraphics[width=0.90\columnwidth]{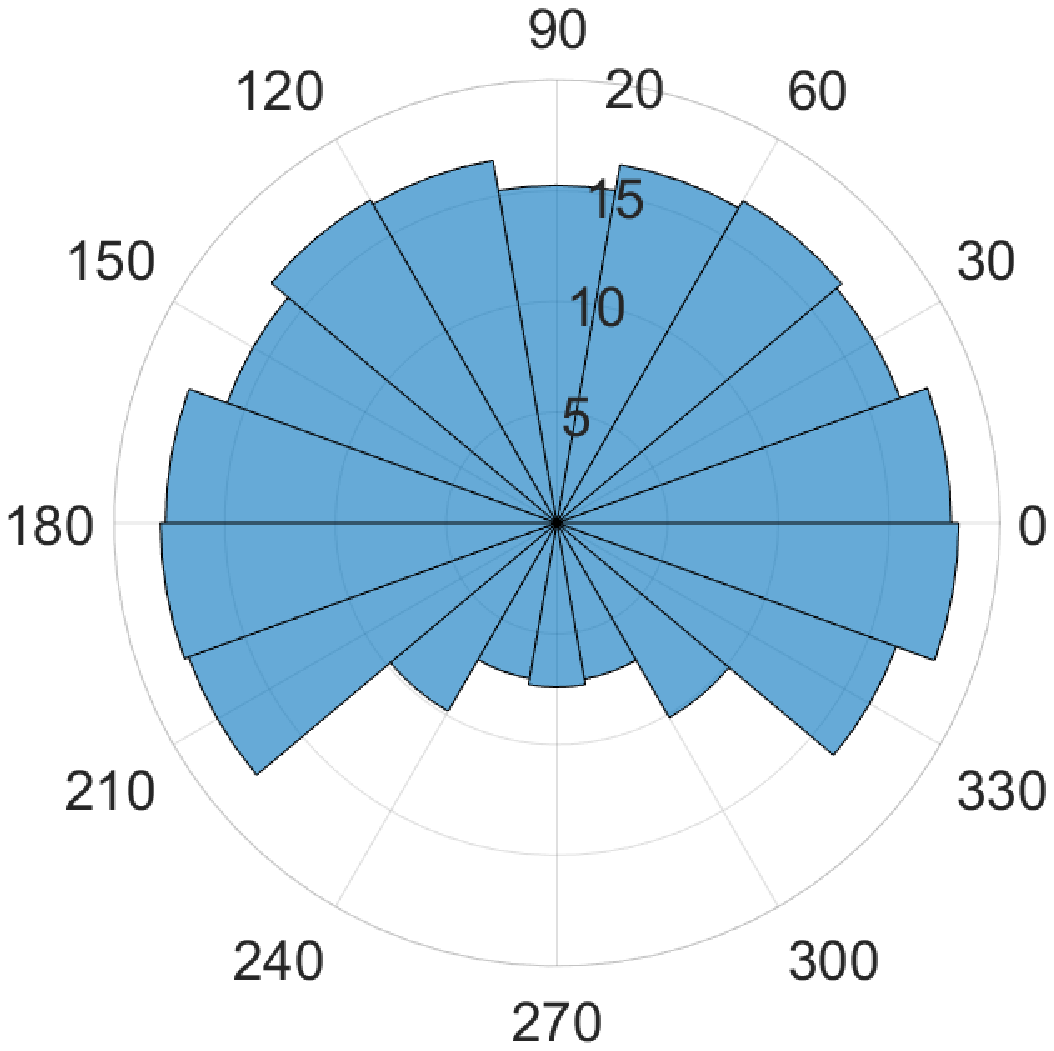}\\
	(b)
	\end{tabular} 
\end{center}
 	\caption{Mean velocity $<\overline{v}>$ as a function of (a) the radial position $r$ and (b) the angle with respect to the transverse direction. Velocities obtained for a dune formed with $m$ = 10.3 g of type 2 grains under $U$ = 0.295 m/s, corresponding to $Re$ = $2.2 \times 10^4$, $Re_*$ = 8 and $\theta$ = 0.038.}
 	\label{fig:hist1}
\end{figure}

Figures \ref{fig:hist1}(a) and \ref{fig:hist1}(b) present the mean velocity $<\overline{v}>$ as functions of the radial position $r$ (with origin at the dune centroid) and the angle with respect to the transverse direction, respectively. In Fig. \ref{fig:hist1}(a) the abscissa corresponds to the radial position, the ordinate to the magnitude of the velocity vector, and the width of bars to the interval between the considered radial positions. In Fig. \ref{fig:hist1}(b), the numbers along the perimeter correspond to angles with respect to the transverse direction (the water flow direction is 270$^{\circ}$) and the height of bars to the magnitude of velocity, which can be measured using the radial scale along the 80$^{\circ}$ line.

Figures \ref{fig:hist1}(a) and \ref{fig:hist1}(b) corroborate the observation made from Fig. \ref{fig:quiver1} about the existence of higher velocities in the centroid region, but now we can obtain readily the magnitude of velocities with respect to the dune centroid. The radial and angular charts for all the other test runs are available as Supplementary Material \cite{Supplemental}. We note from those charts that the mean velocities close to the centroid are around 8-10\% of the water velocity, and around 1-5\% of the water velocity far from the centroid, i.e., at the dune periphery.

It is important to remark here that the mean Eulerian velocities were computed by considering all displacements of tracers, even those which were accelerating or stopping. Such velocity criteria is different from the Langangian mean velocity presented in Subsection \ref{subsection_lagrangian}, where we did not considered strongly accelerating and decelerating tracers.

\subsection{Lagrangian Framework}
\label{subsection_lagrangian}

\begin{figure}[ht]
\begin{center}
	\begin{tabular}{c}
	\includegraphics[width=0.60\columnwidth]{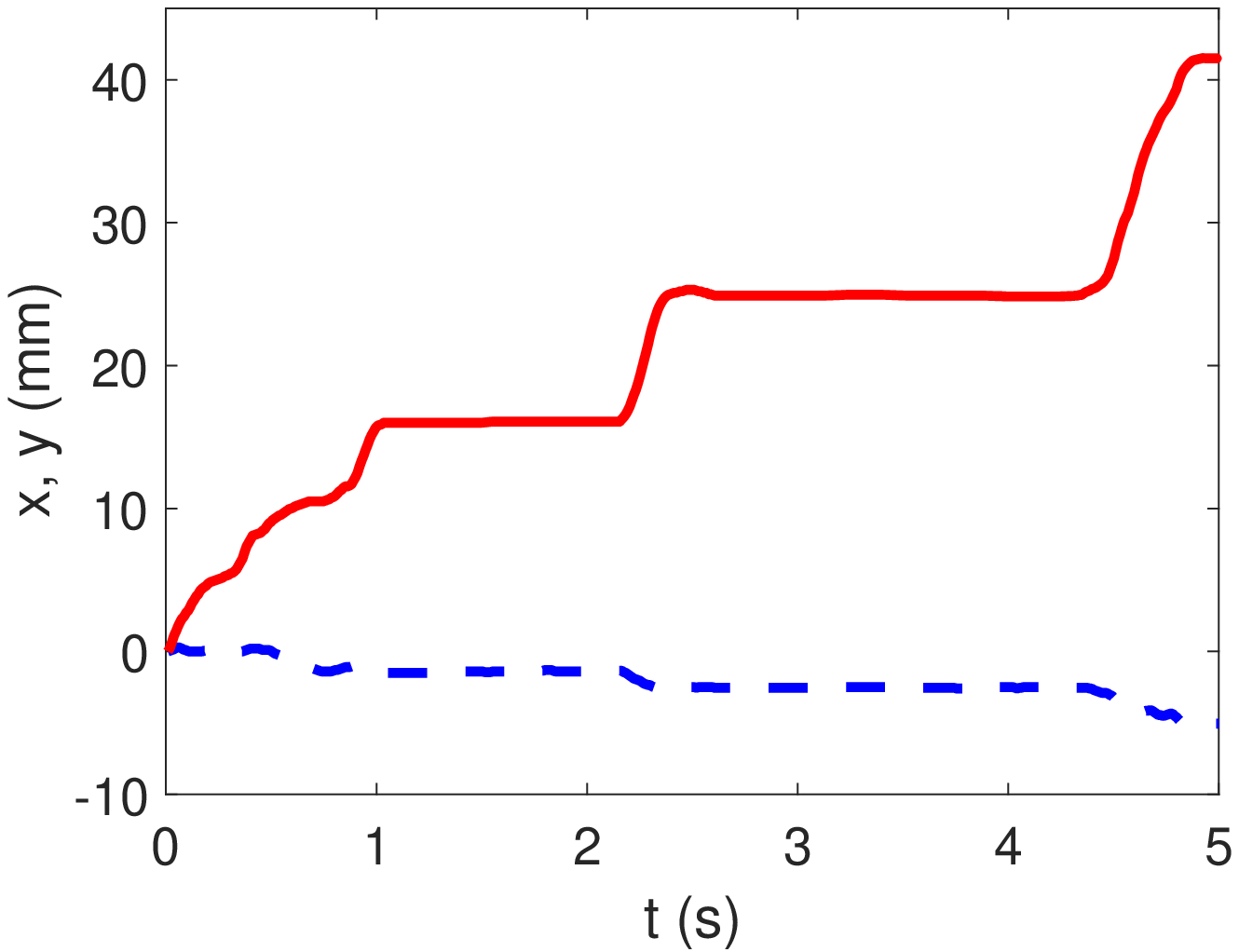}\\
	(a)\\
	\\
	\\
	\includegraphics[width=0.60\columnwidth]{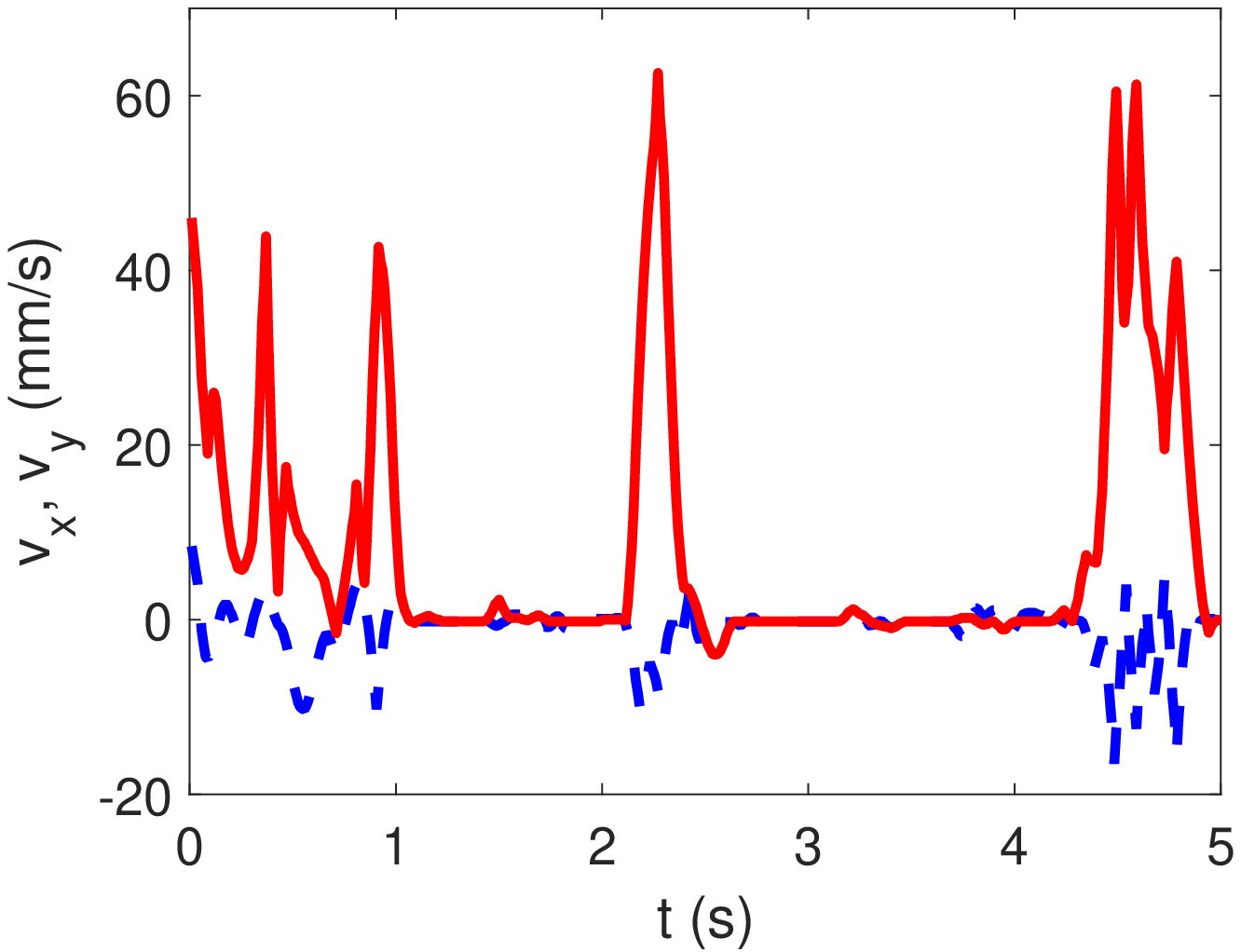}\\
	(b)
	\end{tabular} 
\end{center}
 	\caption{Example of tracking of a single grain over a dune formed with $m$ = 10.3 g of grains of type 1 under $U$ = 0.365 m/s, corresponding to $Re_*$ = 10 and $\theta$ = 0.055. (a) Longitudinal  and transverse positions, $y$ and $x$, and (b) longitudinal and transverse velocities, $v_y$ and $v_x$, of a tracer as functions of time. The red continuous and blue dashed curves correspond to the $y$ and $x$ components, respectively. The origin of the $y(t),x(t)$ graphic is at the original position of the tracked particle, which in this case was to the right of the barchan symmetry line.}
 	\label{fig:lagrangian}
\end{figure}

\begin{figure}[ht]
\begin{center}
	\begin{tabular}{c}
	\includegraphics[width=0.60\columnwidth]{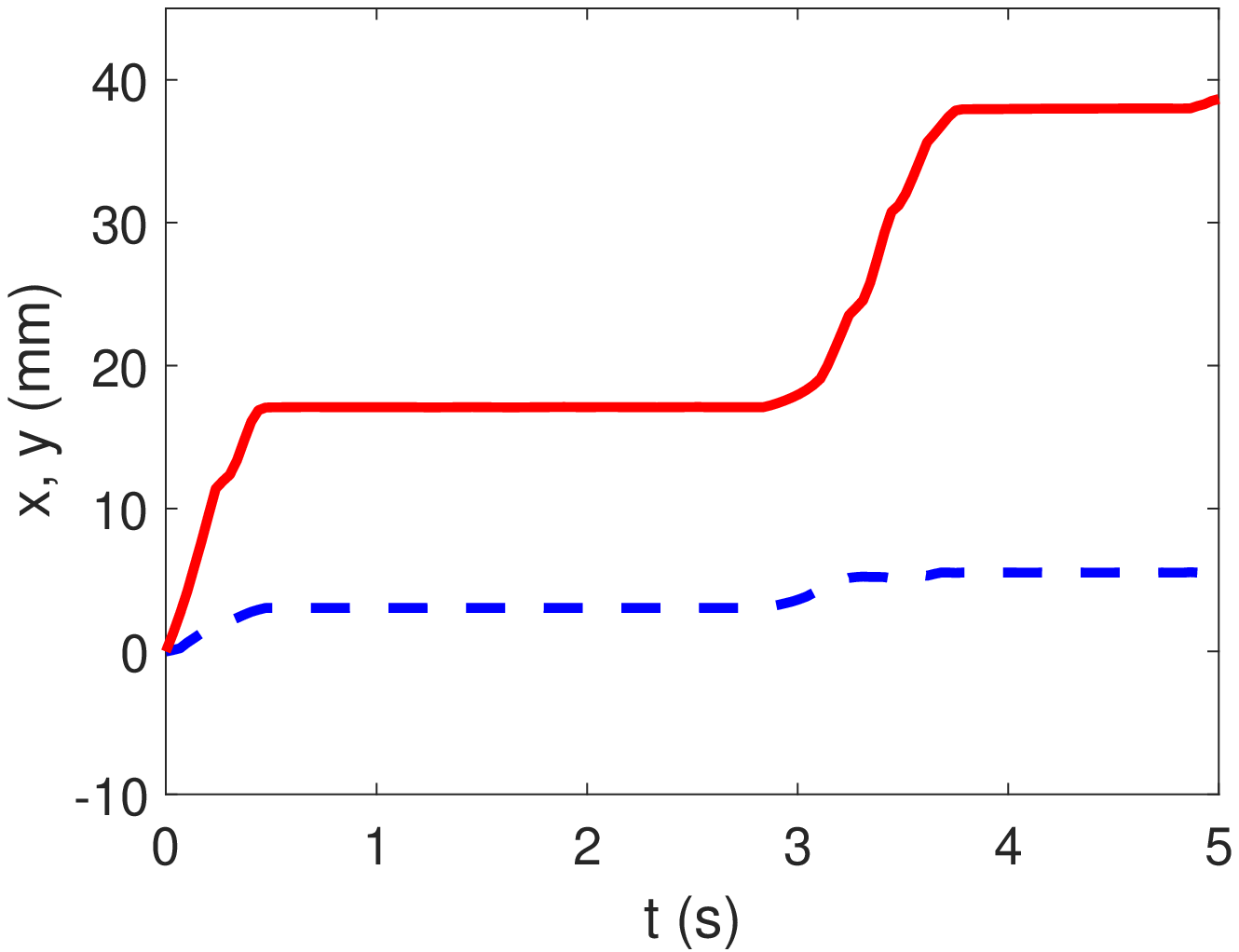}\\
	(a)\\
	\\
	\\
	\includegraphics[width=0.60\columnwidth]{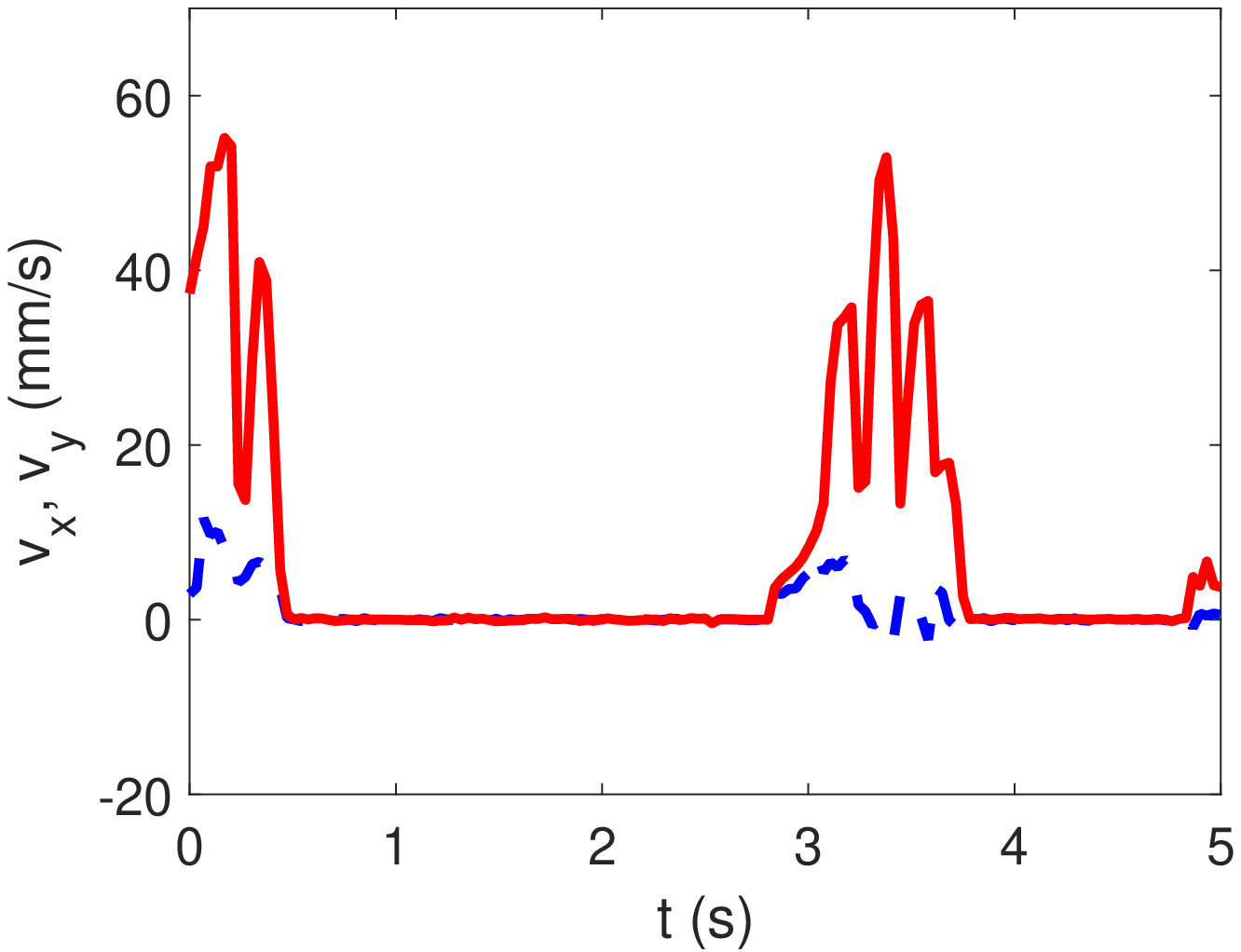}\\
	(b)
	\end{tabular} 
\end{center}
 	\caption{Example of tracking of a single grain over a dune formed with $m$ = 10.3 g of grains of type 1 under $U$ = 0.365 m/s, corresponding to $Re_*$ = 10 and $\theta$ = 0.055. (a) Longitudinal  and transverse positions, $y$ and $x$, and (b) longitudinal and transverse velocities, $v_y$ and $v_x$, of a tracer as functions of time. The red continuous and blue dashed curves correspond to the $y$ and $x$ components, respectively. The origin of the $y(t),x(t)$ graphic is at the original position of the tracked particle, which in this case was to the left of the barchan symmetry line.}
 	\label{fig:lagrangian2}
\end{figure}

We obtained the centroid positions of tracers as functions of time with the procedure described in Subsection \ref{section_image_processing}, and from them we computed the distances and instantaneous velocities for the tracers. Figs. \ref{fig:lagrangian} and \ref{fig:lagrangian2} present examples of trajectories of tracers over a dune formed with $m$ = 10.3 g of type 1 grains under $U$ = 0.365 m/s, which corresponds to $Re_*$ = 10 and $\theta$ = 0.055. Figs. \ref{fig:lagrangian}a and \ref{fig:lagrangian}b show, respectively, the longitudinal and transverse positions, $y$ and $x$, and the longitudinal and transverse velocities, $v_y$ and $v_x$, of a tracer as functions of time. Fig. \ref{fig:lagrangian} corresponds to a tracer that was on the right side (with respect to the flow direction) of the barchan symmetry line. Figs. \ref{fig:lagrangian2}a and \ref{fig:lagrangian2}b show the same thing for a tracer that was on the left side of the barchan symmetry line. In Figs. \ref{fig:lagrangian} and \ref{fig:lagrangian2}, the red continuous and blue dashed curves correspond to the $y$ and $x$ components, respectively.

The behavior depicted in Figs. \ref{fig:lagrangian} and \ref{fig:lagrangian2} is typical of the ensemble of tracers tracked in our tests: they move mainly in the flow direction, with small transverse components, and present an intermittent motion, having periods of acceleration, deceleration and at rest. The intermittent motion is similar to that observed over plane beds for water velocities and grain diameters of the same order of magnitude as in the present study \cite{Penteado}. Transverse movements can be significant close to the dune flanks and downstream the lee side because the water flow direction changes significantly in these regions, as shown by Alvarez and Franklin \cite{Alvarez3} in the case of growing barchans, and as can be remarked in the movie available as Supplementary Material \cite{Supplemental}

For the moving grains, as the ones of Figs. \ref{fig:lagrangian} and \ref{fig:lagrangian2}, we computed displacements and velocities by considering only the moving periods. In Figs. \ref{fig:lagrangian}a and \ref{fig:lagrangian2}a, the moving periods correspond to positive slopes of the continuous line; therefore, our scripts identify those positive slopes for a given tracer, and then compute the displacements as the differences between the initial and final positions and the velocities as the slope values. This procedure gives the displacements between resting times and the corresponding velocities experienced by a single tracer. The scripts repeat the same computations for all the moving tracers of each test run, and afterward compute single averages and standard deviations for the entire test run. Because the mean lagrangian velocity is computed as the average of measured slopes, it does not take into account the accelerating and decelerating stages, being higher than the corresponding Eulerian averages.

\begin{sloppypar}
Table \ref{tab_dim} shows the mean values and standard deviations of longitudinal and transverse displacements and velocities, for each test run. The table presents the initial mass $m$, grain diameter $d$, mean cross-sectional velocity of the fluid $U$, shear velocity $u_*$, mean transverse displacement $\Delta x$, mean longitudinal displacement $\Delta y$, standard deviation of the mean transverse displacement $\sigma_{\Delta x}$, standard deviation of the mean longitudinal displacement $\sigma_{\Delta y}$, mean transverse velocity $<v_{x,lag}>$, mean longitudinal velocity $<v_{y,lag}>$, standard deviation of the mean transverse velocity $\sigma_{v x}$, standard deviation of the mean longitudinal velocity $\sigma_{v y}$, displacement time $t_d$, number of images $N_{im}$, total number of moving tracers $N_{mv}$, and movement density $\rho_{mv}$. The displacement time was computed as $\Delta y / <v_{y,lag}>$, and the shear velocity $u_*$ corresponds to the flow over the channel wall (of acrylic), the latter computed from velocity profiles measured with PIV (Particle Image Velocimetry).
\end{sloppypar}

\begin{table*}[t]
 	\begin{center}
 		\caption{Initial mass $m$, grain diameter $d$, mean cross-sectional fluid velocity $U$, shear velocity $u_*$, mean transverse displacement $\Delta x$, mean longitudinal displacement $\Delta y$, standard deviation of the mean transverse displacement $\sigma_{\Delta x}$, standard deviation of the mean longitudinal displacement $\sigma_{\Delta y}$, mean transverse velocity $<v_{x,lag}>$ , mean longitudinal velocity $<v_{y,lag}>$, standard deviation of the mean transverse velocity $\sigma_{v x}$, standard deviation of the mean longitudinal velocity $\sigma_{v y}$, displacement time $t_d$, number of images $N_{im}$, total number of moving tracers $N_{mv}$, and movement density $\rho_{mv}$.}
 		\begin{tabular}{c c c c c c c c c c c c c c c c}
 			\hline\hline
 			$m$ & $d$ & $U$ & $u_*$& $\Delta_x$ & $\Delta_y$ & $\sigma_{\Delta x}$ & $\sigma_{\Delta y}$ & $<v_{x,lag}>$ & $<v_{y,lag}>$ & $\sigma_{v x}$ & $\sigma_{v y}$ & $t_d$ & $N_{im}$ & $N_{mv}$ & $\rho_{mv}$\\			
			g & mm & m/s & m/s & mm & mm & mm & mm & mm/s & mm/s & mm/s & mm/s & s & $\cdots$ & $\cdots$ & m$^{-2}$\\
 			\hline
 			10.3 & 0.5 & 0.365 & 0.0202 & 0.1 & 17.6 & 4.2 & 15.3 & 0.4 & 71.1 & 18.9 & 37.3 & 0.25 & 999 & 5994 & 217391\\
			10.3 & 0.5 & 0.365 & 0.0202 & 0.3 & 18.0 & 4.5 & 16.4 & 0.9 & 69.7 & 18.7 & 38.2 & 0.26 & 999 & 5373 & 194869\\
			10.3 & 0.5 & 0.365 & 0.0202 & 0.2 & 16.5 & 4.1 & 15.3 & 1.1 & 64.6 & 17.7 & 35.9 & 0.26 & 999 & 6208 & 225153\\
			6.2 & 0.5 & 0.365 & 0.0202 & 0.1 & 15.7 & 3.9 & 12.9 & 0.1 & 69.8 & 19.4 & 36.2 & 0.22 & 1299 & 7400 & 347359\\
			6.2 & 0.5 & 0.365 & 0.0202 & 0.0 & 15.7 & 4.1 & 12.6 & 0.3 & 72.2 & 20.0 & 36.9 & 0.22 & 999 & 7047 & 430125\\
			6.2 & 0.5 & 0.365 & 0.0202 & 0.0 & 17.2 & 4.1 & 15.2 & 0.1 & 70.3 & 18.5 & 37.0 & 0.24 & 1299 & 3410 & 160067\\
			10.3 & 0.5 & 0.295 & 0.0168 & -0.1 & 19.8 & 5.4 & 25.4 & -0.1 & 42.7 & 13.7 & 27.9 & 0.46 & 5997 & 8393 & 50708\\
			10.3 & 0.5 & 0.295 & 0.0168 & 0.3 & 22.2 & 5.8 & 29.1 & 0.6 & 44.4 & 13.8 & 28.5 & 0.50 & 5997 & 8317 & 50249\\
			10.3 & 0.5 & 0.295 & 0.0168 & 0.4 & 21.3 & 5.5 & 26.0 & 0.6 & 44.8 & 13.6 & 38.8 & 0.48 & 5997 & 11480 & 69358\\
			6.2 & 0.5 & 0.295 & 0.0168 & 0.4 & 23.3 & 5.6 & 29.4 & 0.9 & 46.3 & 13.4 & 29.7 & 0.50 & 5997 & 8149 & 82856\\
			6.2 & 0.5 & 0.295 & 0.0168 & 0.3 & 21.9 & 5.5 & 27.0 & 0.4 & 46.5 & 14.2 & 29.8 & 0.47 & 5997 & 7967 & 81006\\
			6.2 & 0.5 & 0.295 & 0.0168 & 0.1 & 21.6 & 5.7 & 28.0 & 0.1 & 45.3 & 13.9 & 29.4 & 0.48 & 5997 & 7447 & 75719\\
			10.3 & 0.5 & 0.243 & 0.0141 & -0.1 & 21.1 & 6.3 & 30.1 & -1.0 & 34.1 & 12.8 & 22.1 & 0.62 & 5997 & 2516 & 15201\\
			10.3 & 0.5 & 0.243 & 0.0141 & 0.4 & 20.0 & 5.6 & 27.9 & 0.8 & 33.5 & 11.9 & 22.7 & 0.60 & 5997 & 2060 & 12446\\
			10.3 & 0.5 & 0.243 & 0.0141 & 0.3 & 20.6 & 5.5 & 28.8 & 0.7 & 34.2 & 12.6 & 23.1 & 0.60 & 5997 & 2495 & 15074\\
			6.2 & 0.5 & 0.243 & 0.0141 & -0.1 & 17.0 & 5.0 & 25.2 & -0.4 & 29.9 & 12.2 & 20.8 & 0.57 & 5997 & 1485 & 15099\\
			6.2 & 0.5 & 0.243 & 0.0141 & -0.2 & 21.5 & 5.8 & 31.9 & -0.2 & 37.0 & 13.2 & 23.4 & 0.58 & 5997 & 1490 & 15150\\
			6.2 & 0.5 & 0.243 & 0.0141 & -0.2 & 21.8 & 5.8 & 34.0 & -0.1 & 31.5 & 13.4 & 21.9 & 0.69 & 5997 & 1205 & 12252\\
			10.3 & 0.2 & 0.295 & 0.0168 & 0.1 & 11.4 & 2.8 & 11.3 & 0.3 & 34.0 & 9.0 & 18.3 & 0.34 & 5997 & 70513 & 170406\\
			10.3 & 0.2 & 0.295 & 0.0168 & 0.1 & 11.1 & 2.7 & 10.5 & 0.2 & 34.6 & 9.3 & 18.3 & 0.32 & 5997 & 78275 & 189165\\
			10.3 & 0.2 & 0.295 & 0.0168 & 0.0 & 10.7 & 2.7 & 10.2 & 0.1 & 33.7 & 9.1 & 17.9 & 0.32 & 5997 & 77985 & 188464\\
			6.2 & 0.2 & 0.295 & 0.0168 & -0.1 & 10.4 & 2.5 & 11.4 & -0.6 & 29.5 & 7.6 & 15.7 & 0.35 & 5997 & 53786 & 218752\\
			6.2 & 0.2 & 0.295 & 0.0168 & -0.1 & 10.9 & 2.7 & 11.5 & -0.2 & 32.0 & 8.5 & 17.4 & 0.34 & 5997 & 55156 & 224324\\
		\end{tabular}
	\label{tab_dim}
\end{center}
\end{table*}

From Tab. \ref{tab_dim}, we observe that, by averaging over the whole dune, grains move mainly in the longitudinal direction, with both $\Delta_y / \Delta_x$ and $<v_{y,lag}> / <v_{x,lag}>$ having an order of magnitude equal to 100. It is important to note that, if we consider individual grains migrating from and to certain parts of the dune, the transverse components can be significant. As an example, that happens to the grains that move along the periphery of the dune, i.e., that go around the dune from its leading edge to one of its horns, as shown by Alvarez and Franklin \cite{Alvarez3}. In the present study, we investigate displacements averaged over the whole barchan. For these averages, as the fluid shearing increases, the longitudinal displacement decreases slightly while the longitudinal velocity increases and the displacement time decreases. This result is intriguing and was not \textit{a priori} expected. However, by analyzing the movies, we can observe that grains indeed travel shorter distances faster for higher shear velocities, but the quantity of moving grains is much higher, which indicates higher transport rates. The total number of moving tracers identified in all images $N_{mv}$, irrespective if they were the same from previous images (i.e., they could be already moving in previous images), are also shown in Tab. \ref{tab_dim}. A measure of the density of moving grains can be computed by considering the total field of each test, the total number of images of each test run $N_{im}$, and that the behavior of tracers is characteristic of the ensemble of grains. With these assumptions, the following quantity is proportional to the density of moving grains,

\begin{equation}
\rho_{mv} \,=\, \frac{50N_{mv}}{N_{im}A_{d}} \,,
\label{eq_density}
\end{equation}

\noindent where $A_{d}$ is the characteristic dune area (its volume divided by 10$d$) and $\rho_{mv}$ is called here movement density because it is not exactly the density of moving grains, but it is a reasonable approximation of it. Table \ref{tab_dim} shows that, indeed, the density of moving grains increases with the fluid shearing. Comparisons with Penteado and Franklin \cite{Penteado} show densities of the same order of magnitude for lower Shields numbers, and higher for higher Shields numbers (please see Tab. \ref{tab.1} for the Shields numbers). The computations employed for obtaining $\rho_{mv}$ in Penteado and Franklin \cite{Penteado} were different from ours; however, even considering uncertainties and errors, the results point to an increase in $\rho_{mv}$ over a barchan dune when compared to a flat bed, at least for Shields numbers equal or greater than 0.055. The higher density of moving grains is in accordance with the conclusions of Lajeunesse et al. \cite{Lajeunesse}, who found that the erosion rate depends locally on the shear stress caused by the liquid. Concerning variations with the grain diameter, both the longitudinal displacement and velocity increase with its augmentation.

Two important length scales for bed load are the grain diameter $d$ and the saturation length $L_{sat}\,\sim\,L_{drag}$, the latter being a length scale for the stabilization of sand flux after or downstream changes in fluid or sediment conditions \cite{Andreotti_1,Andreotti_2,Hersen_1}. One way to investigate if they are pertinent scales is by normalizing the longitudinal displacements by both parameters; therefore, we computed the dimensionless displacements $\Delta_{y,d}^{ad}\,=\,\Delta_y / d$ and $\Delta_{y,Ldrag}^{ad}\,=\,\Delta_y / L_{drag}$. Concerning the velocities, two important scales are the mean and shear velocities, $U$ and $u_*$, respectively. However, instead of using directly the shear velocity $u_*$ as a scale, the use the mean shear at the grain scale, $\partial_y u\,d$ = $u_*^2 d / \nu$, where $\partial_y u$ is the vertical gradient of the fluid velocity, may be more convenient. This scale represents the mean velocity at the grain scale. Therefore, we computed the dimensionless velocities $V_{y,U}^{ad} \,=\, <v_{y,lag}> / U$ and $V_{y,*}^{ad} \,=\, <v_{y,lag}> / ( u_*^2 d / \nu )$. Finally, a natural time scale for bed load is the settling time, $d / U_s$, where $U_s$ is the settling velocity of one single grain. This time scale represents the time that a grain takes to fall a distance equivalent to its diameter. Therefore, we computed the dimensionless displacement times $t_d^{ad} \,=\, t_d U_s/d$. 

\begin{sloppypar}
Table \ref{tab.1} shows the average values of equivalent test runs.  It presents the initial mass $m$, grain diameter $d$, Reynolds number based on the mean velocity and hydraulic diameter $Re$, Reynolds number based on the shear velocity and grain diameter $Re_{*}$, Shields number $\theta$, mean longitudinal displacement normalized by the grain diameter $\Delta_{y,d}^{ad}$, mean longitudinal displacement normalized by the drag length $\Delta_{y,Ldrag}^{ad}$, longitudinal velocity normalized by the cross-sectional mean velocity $V_{y,U}^{ad}$, longitudinal velocity normalized by the mean shear and grain diameter $V_{y,*}^{ad}$, and displacement time normalized by the settling velocity and grain diameter $t_d^{ad}$. In Tab. \ref{tab.1}, the dimensional values were averaged over the test runs under the same experimental conditions; therefore, it corresponds to a summary of all experiments.
\end{sloppypar}
 
 \begin{table*}[t]
 	\begin{center}
 		\caption{Initial mass $m$, grain diameter $d$, Reynolds number based on the mean velocity and hydraulic diameter $Re$, Reynolds number based on the shear velocity and grain diameter $Re_{*}$, Shields number $\theta$, mean longitudinal displacement normalized by the grain diameter $\Delta_{y,d}^{ad}$, mean longitudinal displacement normalized by the drag length $\Delta_{y,Ldrag}^{ad}$, longitudinal velocity normalized by the cross-sectional mean velocity $V_{y,U}^{ad}$, longitudinal velocity normalized by the mean shear and grain diameter $V_{y,*}^{ad}$, and displacement time normalized by the settling velocity and grain diameter $t_d^{ad}$. Averaged values for test runs corresponding to the same experimental conditions.}
 		\begin{tabular}{c c c c c c c c c c}
 			\hline\hline
 			$m$ & $d$ & $Re$ & $Re_*$ & $\theta$ & $\Delta_{y,d}^{ad}$ & $\Delta_{y,Ldrag}^{ad}$ &  $V_{y,U}^{ad}$ &  $V_{y,*}^{ad}$ & $t_d^{ad}$\\			
			g & mm & $\cdots$ & $\cdots$ & $\cdots$ & $\cdots$ & $\cdots$ & $\cdots$ & $\cdots$ &$ \cdots$\\
 			\hline
 			10.3 & 0.5 & 2.8 $\times$ 10$^4$ & 10 & 0.055 & 34.7 & 13.9 & 0.19 & 0.34 & 37.1\\
 			6.2 & 0.5 & 2.8 $\times$ 10$^4$ & 10 & 0.055 & 32.4 & 13.0 & 0.19 & 0.35 & 33.6\\
 			10.3 & 0.5 & 2.2 $\times$ 10$^4$ & 8 & 0.038 & 42.2 & 16.9 & 0.15 & 0.31 & 70.4\\
			6.2 & 0.5 & 2.2 $\times$ 10$^4$ & 8 & 0.038 & 44.5 & 17.8 & 0.16 & 0.33 & 71.0\\
			10.3 & 0.5 & 1.9 $\times$ 10$^4$ & 7 & 0.027 & 41.1 & 16.4 & 0.14 & 0.34 & 90.0\\
			6.2 & 0.5 & 1.9 $\times$ 10$^4$ & 7 & 0.027 & 40.2 & 16.1 & 0.13 & 0.33 & 90.1\\
			10.3 & 0.2 & 2.2 $\times$ 10$^4$ & 3 & 0.095 & 55.3 & 22.1 & 0.12 & 0.61 & 37.2\\
			6.2 & 0.2 & 2.2 $\times$ 10$^4$ & 3 & 0.095 & 53.3 & 21.3 & 0.10 & 0.55 & 39.7\\			
 		\end{tabular}
 		\label{tab.1}
 	\end{center}
 \end{table*}

For the ensemble of our tests, the mean displacement decreases slightly with the Shields number, indicating a decrease in traveled distances with the fluid shear. By considering that the shear velocities over a barchan dune reach, close to its crest, 1.4 of the value over the flat bed \cite{Andreotti_1,Andreotti_2}, Shields numbers over barchans reach values of approximately twice those over flat beds. The values of $u_*$ reported in the present paper were computed from profiles measured over the channel wall; however, $u_*$ values measured over plane beds with similar grains showed that the flow is close to hydraulic smooth regimes and the feedback effect is relatively weak \cite{Franklin_9}. Therefore, by considering the values of $2 \theta$ as representative of the normalized shear over the barchan, and by observing that displacement lengths over the barchan dune are between 13 and 18 $L_{drag}$ and 30 and 40 $d$ for type 1 grains and around 21-22 $L_{drag}$ and 55 $d$ for type 2, the values of $\Delta_{y,d}^{ad}$ can be directly compared with values reported by Lajeunesse et al. \cite{Lajeunesse} and Penteado and Franklin \cite{Penteado} for plane beds with the same range of Shields values. The comparison shows that mean values over barchan dunes are three to five times higher than those obtained over liquid-sheared plane beds. Considering that the threshold Shields $\theta_c$ is higher over the dune positive slope and that the comparison is made for the same range of Shields values, the longer displacements over the barchan dune indicates that details of the structure of the water flow over the dune affect substantially bed load characteristics. It has been proposed that the curvature of streamlines induce higher turbulent stresses at low regions of the boundary layer close to the dune leading edge and smaller ones at low regions close to the dune crest (first proposed by Wiggs et al. \cite{Wiggs} and afterward measured experimentally, for example, for 2D dunes by C\'u\~nez et al. \cite{Franklin_10}). The higher turbulent stresses over the upwind face of barchans would then be responsible for the longer distances traveled by grains, but this rests to be investigated further.

\begin{sloppypar}
The displacement velocities increase with the cross-sectional mean velocity of the fluid, varying roughly between  10 and 20 \% of $U$ for type 1 grains. When considering the grain diameter at the same $U$, the value of $V_{y,U}^{ad}$ decreases from approximately 0.15 for type 1 grains to approximately 0.10 for type 2. In terms of the mean shear at the grain scale, the displacement velocity remains at approximately 30 \% of $u_*^2 d / \nu$ for type 1 grains, at all flow conditions, and reaches approximately 60 \% of $u_*^2 d / \nu$ for type 2 grains. The values for type 1 grains are equal to those found by Penteado and Franklin \cite{Penteado} for grains of same mean diameter. This indicates that, although the fluid structure changes over the dune \cite{Wiggs}, causing displacements up to five times longer, the velocity of grains remains roughly the same as that over the flat bed. From the dimensional values, type 2 grains have displacement velocities that are smaller than that of larger grains; therefore, comparisons using $V_{y,*}^{ad}$ show that smaller grains move slower over the dune because they are exposed to lower mean velocities.
\end{sloppypar}

\begin{figure}[ht]
\begin{center}
	\begin{tabular}{c}
	\includegraphics[width=0.85\columnwidth]{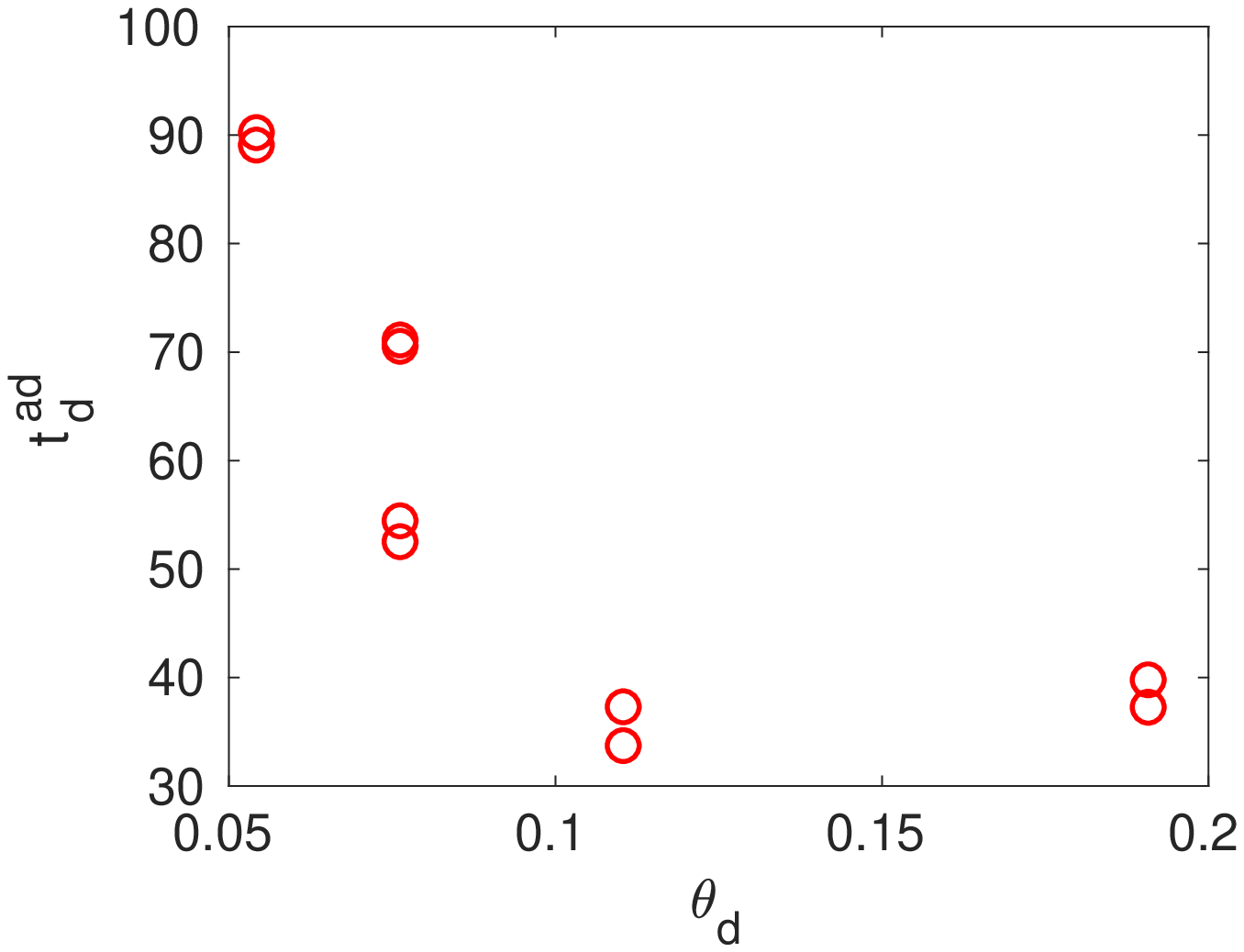}\\
	(a)\\
	\includegraphics[width=0.85\columnwidth]{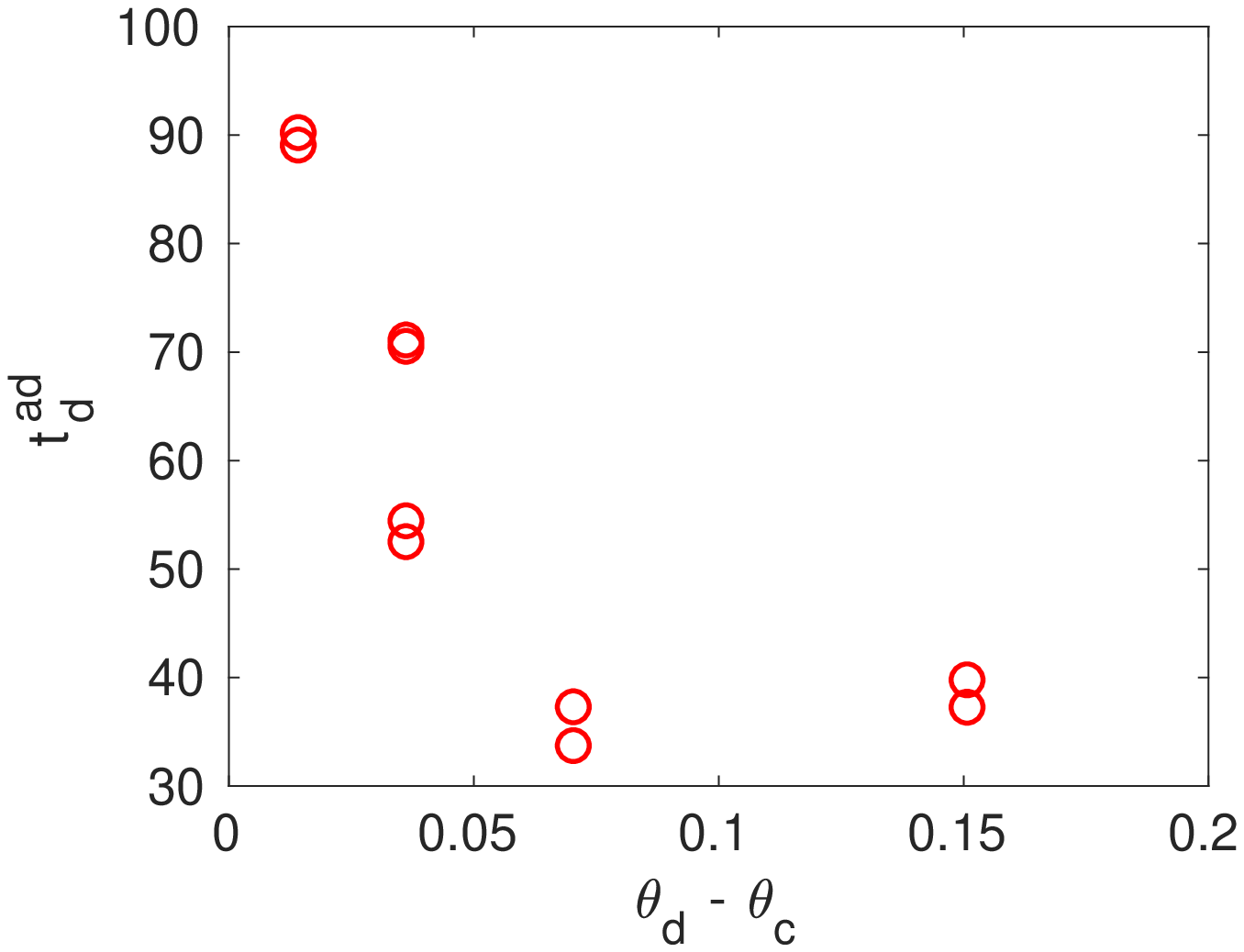}\\
	(b)
	\end{tabular} 
\end{center}
 	\caption{Normalized displacement time $t_d^{ad}$ as functions of (a) the Shields number over the dune $\theta_d$ and (b) the excess of Shields number over the dune, $\theta_d \, - \, \theta_c$}
 	\label{fig:td_theta}
\end{figure}

Finally, we observe that the displacement time decreases with the shear stresses caused by the fluid and increases with the grain diameter, varying within roughly 30 and 90 \% of the settling time for type 1 grains and being approximately 40 \% of the settling time for type 2 grains. At first sight, the displacement time seems to not scale with the Shields number. However, by considering that most of dunes were close to incipient bed-load conditions, the type 1 grains may be too close to incipient conditions at the lower fluid velocities. Figures \ref{fig:td_theta}a and \ref{fig:td_theta}b present $t_d^{ad}$ as functions of $\theta_d$ and of $\theta_d - \theta_c$, respectively, where $\theta_d \,=\, 2\theta$ and $\theta_c$ was considered here as 0.04 \cite{Yalin_2,Buffington_1}. The figure shows a negative variation for $\theta_d < 0.1$ and a constant value for $\theta_d \geq 0.1$, indicating the presence two distinct asymptotic behaviors: one close to incipient conditions and other farther from it. However, given the relatively narrow variations of the pertinent parameters in the present study, the existence of different behaviors for $t_d^{ad}$ close and far from bed-load inception needs to be investigated further.

\section{Conclusions}

This paper presented an experimental investigation of the displacements of individual grains over subaqueous barchan dunes. Granular piles were placed in a closed conduit of transparent material where a turbulent water flow was imposed afterward. With this procedure, a single barchan dune was formed, which was filmed with a high-speed camera as it migrated along the channel. The velocity fields and the trajectories of individual grains were computed from the acquired images with numerical scripts written in the course of this work.

For the Eulerian framework, the instantaneous fields of bed-load velocity were obtained for each image by subtracting the positions of moving tracers (grains with different color) from the previous image. Time-averaged fields were then computed by generating a Cartesian mesh with origin at the dune centroid, computing the mean value in each cell, and computing a time average for each cell. These fields are available for all the test runs as Supplementary Material \cite{Supplemental}. The use of graphics of the magnitude of mean velocity as functions of the radial position $r$ (with origin at the dune centroid) and the angle with respect to the transverse direction allows the easy identification of the velocities magnitude and their respective positions over the dune. Depending on the region over the barchan, we found that the mean velocity of grains varies roughly within 1 and 10\% of the cross-sectional mean velocity of the fluid, with the highest velocities in the centroid region, and with important transverse components close to the lateral flanks of the dune and in avalanche and recirculation regions.

For the Lagrangian framework, we observed that individual grains move mainly in the flow direction, with small transverse components, and presenting an intermittent motion, with periods of acceleration, deceleration and at rest. For the range of diameters employed in the present study, our results show that, with the increase of the fluid shearing, the dimensional form of the longitudinal displacement decreases while the longitudinal velocity increases, which was not necessarily expected \textit{a priori}. By computing an average over the whole barchan, we found that the average displacement of grains varies within 30 and 60 grain diameters, which is three to five times higher than displacements obtained over liquid-sheared plane beds by Lajeunesse et al. \cite{Lajeunesse} and Penteado and Franklin \cite{Penteado}. One possible explanation for higher values over the barchan dune is the presence higher turbulent stresses over the upwind face of barchans, as proposed by Wiggs et al. \cite{Wiggs} and measured by C\'u\~nez et al. \cite{Franklin_10}, that would be responsible for the longer distances traveled by grains, but this rests to be investigated further. We found that the average velocity varies within 10 and 20 \% of the cross-sectional mean velocity of the fluid and 30 and 60 \% of the velocity based on the mean shear at the grain scale. The values for type 1 grains are equal to those found by Penteado and Franklin \cite{Penteado} for a flat bed, pointing out that changes in the fluid structure over the dune \cite{Wiggs} do not affect significantly the velocity of grains.

Our results showed that the displacement time varies between 30 and 90 \% of the settling time, decreasing with the Shields number for $\theta_d < 0.1$ and remaining constant for $\theta_d \geq 0.1$. This indicates the existence of two distinct asymptotic behaviors, one close to incipient bed load and other farther from it. However, this needs to be investigated further.

Those findings are a contribution to our knowledge on the dynamics of barchan dunes, and can bring significant advancements on numerical simulations involving barchans.

\begin{acknowledgements}
\begin{sloppypar}
Erick Franklin is grateful to FAPESP (grant no. 2016/13474-9), to CNPq (grant no. 400284/2016-2) and to FAEPEX/UNICAMP (grant nos. 2210/18 and 2112/19) for the financial support provided.
\end{sloppypar}
\end{acknowledgements}

\section*{Compliance with ethical standards}
\begin{sloppypar}
\noindent \textbf{Funding} This study was funded by the Funda\c{c}\~ao de Amparo \`a Pesquisa do Estado de S\~ao Paulo - FAPESP (grant no. 2016/13474-9),  Conselho Nacional de Desenvolvimento Cient\'ifico e Tecnol\'ogico - CNPq (grant no. 400284/2016-2) and Fundo de Apoio ao Ensino, Pesquisa e Extens\~ao da Unicamp - FAEPEX/UNICAMP (grant nos. 2210/18 and 2112/19).\\

\noindent \textbf{Conflict of interest} The authors declare that they have no conflict of interest.
\end{sloppypar}


\bibliography{references}
\bibliographystyle{spphys}

\end{document}